\documentclass[reprint,aps,prl,superscriptaddress,floatfix]{revtex4-2}

\usepackage[T1]{fontenc}
\usepackage{graphicx}
\usepackage{amsmath,amssymb}
\usepackage{bm}
\usepackage{orcidlink}
\hypersetup{hidelinks}
\usepackage{xcolor}

\newcommand{\Nmin}{\mathcal{N}_{\min}}
\newcommand{\N}{\mathcal{N}}
\newcommand{\tr}{\operatorname{Tr}}
\newcommand{\Hpair}{H_{\mathrm{pair}}}
\newcommand{\id}{\mathbb{I}}
\newcommand{\runin}[1]{\textit{#1}---} 

\makeatletter
\renewcommand{\p@section}{}
\makeatother

\begin{document}

\title{Exact operator bounds on Leggett--Garg violations}

\author{Rathindra Nath Das\,\orcidlink{0000-0002-4766-7705}}
\email{rathin@mit.edu}
\affiliation{MIT Center for Theoretical Physics---a Leinweber Institute, Massachusetts Institute of Technology, Cambridge, MA 02139, USA}
\affiliation{Institute for Theoretical Physics and Astrophysics and W\"urzburg-Dresden Cluster of Excellence ctd.qmat, Julius-Maximilians-Universit\"at W\"urzburg, Am Hubland, 97074 W\"urzburg, Germany}
\affiliation{Department of Particle Physics and Astrophysics, Weizmann Institute of Science, Rehovot 7610001, Israel}

\author{Arnab Kundu\,\orcidlink{0000-0002-1994-3346}}
\email{arnab.kundu@saha.ac.in}
\affiliation{Theory division, Saha Institute of Nuclear Physics, A CI of Homi Bhabha National Institute, 1/AF, Bidhannagar, Kolkata 700064, India}

\author{Nemai Chandra Sarkar\,\orcidlink{0009-0009-0275-7705}}
\email{nemaichandra.sarkar@saha.ac.in}
\affiliation{Theory division, Saha Institute of Nuclear Physics, A CI of Homi Bhabha National Institute, 1/AF, Bidhannagar, Kolkata 700064, India}

\date{\today}

\begin{abstract}
Leggett--Garg (LG) inequalities test whether correlations measured at different times admit a positive joint probability distribution over histories. A violation requires negative values in any joint quasiprobability distribution reproducing those correlations, but such distributions can have negative values even without a violation. Spacetime density kernels extend the density-matrix description to multiple times and are used to study timelike entanglement. We establish a universal bound that relates LG violations to their negative eigenvalue weight. For LG tests involving any number of measurements, we determine the least compatible kernel negativity through an exact relation valid for any system dimension, initial state, dynamics, or measurement rule. This minimum is nonzero if and only if at least one LG inequality is violated. Three-time L\"uders LG tests certify at most one eighth of negativity, whereas measurements resolving individual spin levels can approach one half. Including additional incompatible qubit observables gives a further exact minimum that can exceed the LG-only ceiling without increasing the LG violation.
\end{abstract}

\maketitle

\runin{Introduction} For spatially separated quantum degrees of freedom, Bell inequalities test whether their correlations admit a local classical description~\cite{Bell:1964kc}. Correlations between different time slices, on the other hand, require a related but conceptually distinct notion of Leggett--Garg inequalities (LGIs)~\cite{Leggett1985-of,Emary:2013wfl,Vitagliano:2022wlr}. A classical description assigns a single positive probability distribution to histories, each specifying the values of an observable at the measurement times. If measurements reveal these values without changing later ones, the measured pair correlations must follow from that probability distribution. Quantum systems can violate the LGIs. Any joint probability description that reproduces the pair correlations must then contain negative weights. Thus, the precise quantitative relation between this negativity and LGI violation is a natural and important question.

Spacetime density kernels put this question in an operator setting~\cite{Milekhin:2025ycm,Das:2026ifj,Das:2025fcd}. They assign a copy of the system's state space to each time and collect multitime correlations in one operator, much as a density matrix describes correlations between different parts of a system~\cite{Fitzsimons:2013gga,Horsman:2017nqa,Fullwood:2022rjd,Das:2025fcd}. This approach also connects to studies of timelike entanglement~\cite{Milekhin:2025ycm,Guo:2025dtq,Das:2026ifj}. Unlike an ordinary density matrix, a Hermitian spacetime kernel can have negative eigenvalues. Their total magnitude defines its negativity~\cite{Das:2026ifj}. 

Negativity alone does not establish an LG violation. Temporal quasiprobabilities can be negative when all LG inequalities hold~\cite{Halliwell_2016,Solinas:2024bjx}. The minimum negativity of a joint quasiprobability distribution has previously been related to LG violations.~\cite{deBarros2016,Meng2016, Meng2017Erratum}. Here, we provide a precise spacetime density kernel description of these violations and distinguish between two questions. We ask how much negativity is required by the LG correlations in every compatible kernel, and how much additional negativity is necessary when additional temporal averages are retained. The distinction is important because a standard LG test fixes only pair correlations, not the single-time averages, higher correlations or other observables. 

We derive bounds on LG violations that every Hermitian, unit-trace density kernel reproducing the measured pair correlations must obey. For measurements at three times, we construct a kernel that attains the least negativity and obtain an exact necessary-and-sufficient relation for LG violation. For these tests, L\"uders measurements, which preserve coherence between states giving the same outcome, certify at most one eighth of negativity. Measurements that resolve these states individually, corresponding to a von Neumann-type update, can exceed this ceiling already for spin one and approach one half in a higher-spin family. We extend the exact minimum and its necessary-and-sufficient condition for LG cycles involving any number of measurement times, for arbitrary system dimension, state, dynamics and measurement rule. Finally, we show that measuring a qubit along two perpendicular directions yields an additional exact minimum without increasing the LG violation.

\runin{Exact bounds on LG violations} We measure an observable $Q$ with outcomes $q=\pm1$ at three ordered times. To determine pair correlations, we repeat the same preparation and measure only at the two times in that pair. The measured pair probabilities, $p_{ij}$ give
\begin{equation}
C_{ij}=\sum_{q_i,q_j=\pm1}q_iq_j\,p_{ij}(q_i,q_j),
\qquad |C_{ij}|\leq1.
\label{eq:Cij}
\end{equation}
There are four LG combinations~\cite{Emary:2013wfl,Mawby:2021mdc}
\begin{equation}
\begin{aligned}
K_1&=C_{12}+C_{23}-C_{13},&
K_2&=C_{12}-C_{23}+C_{13},\\
K_3&=-C_{12}+C_{23}+C_{13},&
K_4&=-C_{12}-C_{23}-C_{13}.
\end{aligned}
\label{eq:Ks}
\end{equation}
Each has a classical upper bound of one. For a definite history, each combination equals either one or minus three. For $K_1$, minus three occurs only for the two histories $(+,-,+)$ and $(-,+,-)$. A violation therefore requires their combined weight to be negative~\cite{Halliwell_2016}. Supplementary Sections~S1 and S3 give the operational definitions and the history-weight calculation.

 A three-time spacetime kernel acts on three copies of a state space, one for each measurement time. It returns multitime correlations through
\begin{equation}
\langle O_1(t_1)O_2(t_2)O_3(t_3)\rangle
=\tr\!\left[H(O_1\otimes O_2\otimes O_3)\right].
\label{eq:kernel}
\end{equation}
The correlation on the left is specified by the chosen measurement prescription. An identity on a leg means no measurement at that time. For our bounds, we require only that $H$ be Hermitian, have unit trace and satisfy $\tr(HQ_iQ_j)=C_{ij}$. Here $Q_i$ is $Q$ on leg $i$ and the identity on the others. We call such a kernel \textit{compatible}. Its negativity is
\begin{equation}
\N(H)=\sum_{\lambda<0}|\lambda|
=\frac{\|H\|_1-1}{2}.
\label{eq:negativity}
\end{equation}
A positive compatible kernel yields ordinary history probabilities when $Q$ is measured on its legs and hence satisfies every LG inequality. The negativity also equals its trace distance to the nearest positive, unit-trace operator~\cite{Das:2026ifj,SM}.

 To bound a violation, associate each LG combination with a witness operator. For the first combination,
\begin{equation}
W_1=Q_1Q_2+Q_2Q_3-Q_1Q_3,
\qquad K_1=\tr(HW_1).
\label{eq:witness}
\end{equation}
The leg observables commute, so $W_1$ has the two history values one and minus three as eigenvalues. Reversing outcome labels on one leg gives the other witnesses with the same spectrum. Write $H=H_+-H_-$ for its positive and negative parts. Since $\tr H_-=\N(H)$ and $\tr H_+=1+\N(H)$,
\begin{equation}
\begin{aligned}
K_\alpha
&=\tr(H_+W_\alpha)-\tr(H_-W_\alpha)\\
&\leq\tr H_++3\tr H_-=1+4\N(H).
\end{aligned}
\label{eq:bound}
\end{equation}
The bound holds for arbitrary leg dimensions, regardless of the dynamics or the measurement rule.

For any three measured pair correlations, we construct a compatible kernel with the least possible negativity. We introduce a two-dimensional space at each time to record the outcomes, with $Q|q\rangle=q|q\rangle$. These spaces need not coincide with the physical state space. We define
\begin{equation}
\begin{aligned}
\Hpair=\frac18\Big[&\id^{\otimes3}
+C_{12}(Q\otimes Q\otimes\id)\\
&+C_{23}(\id\otimes Q\otimes Q)
+C_{13}(Q\otimes\id\otimes Q)\Big].
\end{aligned}
\label{eq:Hpair}
\end{equation}
This kernel reproduces the measured correlation pairs and sets the unspecified single-time averages and three-time correlation to zero. This choice makes no assumption about the physical initial state.

The product basis states $|q_1,q_2,q_3\rangle =|q_1\rangle\otimes|q_2\rangle\otimes|q_3\rangle$, with $q_i=\pm1$, label the eight possible outcome sequences and are eigenstates of $\Hpair$. Reversing all three outcomes leaves their pair products unchanged, so the corresponding eigenvalues are equal. The eight eigenvalues therefore form four pairs,
\begin{equation}
\lambda_\alpha=\frac{1-K_\alpha}{8},
\qquad \alpha=1,\ldots,4,
\quad\text{each twice}.
\label{eq:spectrum}
\end{equation}
The four LG inequalities are exactly its positivity conditions. At most one is violated, since the sum of two distinct combinations is at most two. Thus
\begin{equation}
\begin{aligned}
\Nmin&\equiv\min_{H\ \mathrm{compatible}}\N(H)\\
&=\N(\Hpair)=\frac14\big[\max_\alpha K_\alpha-1\big]_+,
\end{aligned}
\label{eq:main}
\end{equation}
where $[x]_+=\max(x,0)$. The least compatible negativity is nonzero if and only if an LG inequality is violated. The pair kernel attains it for any system, initial state, dynamics or measurement rule. Those choices can change the correlations and hence the numerical minimum, but not the exact relation. The construction also works on the physical time legs of arbitrary dimension and outcome degeneracy; see Supplementary Section~S3.

The three correlations define a point in a cube. The correlations that can be reproduced by a nonnegative joint probability distribution over the outcomes at all three times form a tetrahedron bounded by the four LG inequalities~\cite{Emary:2013wfl}. Outside it, one inequality is violated and the minimum negativity grows in proportion to the violation. Its largest possible value is one half. Figure~\ref{fig:map} shows a two-dimensional view and several measurement protocols; the projection is explained in Supplemental Section~S4.

The minimum negativity concerns only the measured pair correlations. A kernel reproducing more temporal correlations can require additional negativity. To make this distinction precise, we now consider sequential L\"uders measurements.

The full L\"uders kernel $H^{\mathrm L}$ describes measurements of different observables on the physical system. For an initial state $\rho_0$, it is defined by
\begin{equation}
\begin{aligned}
&\tr\!\left[
H^{\mathrm L}(O_1\otimes O_2\otimes O_3)
\right]\\
&\qquad=
\frac14\tr\!\left[
\rho_0\{\{O_3(t_3),O_2(t_2)\},O_1(t_1)\}
\right],
\end{aligned}
\label{eq:HLdefinition}
\end{equation}
where the braces denote anticommutators and $O_i(t_i)$ are Heisenberg operators~\cite{Das:2025fcd}. The relation for every choice of the three observables fixes $H^{\mathrm L}$ uniquely. Inserting identities gives the single-time and pair correlations. For a qubit with Pauli observables, this is the pseudo-density matrix~\cite{Fitzsimons:2013gga}. Supplementary Section~S2 shows the construction.

If we retain only measurements of $Q$, the resulting kernel
is
\begin{equation}
\begin{aligned}
H_Q={}&\Hpair+\frac18\Big[
m_1(Q\otimes\id\otimes\id)+m_2(\id\otimes Q\otimes\id)
\\
&+m_3(\id\otimes\id\otimes Q)+T(Q\otimes Q\otimes Q)\Big],
\end{aligned}
\label{eq:HQ}
\end{equation}
where
\begin{equation}
m_i=\tr(H^{\mathrm L}Q_i),
\qquad
T=\tr(H^{\mathrm L}Q_1Q_2Q_3).
\label{eq:HQmoments}
\end{equation}
Each single-time average is measured without measurements at the other times, whereas $T$ is the average product of the outcomes when all three measurements are performed. The pair correlations are still obtained from their separate two-time runs.

As in $\Hpair$, the tensor products in Eq.~\eqref{eq:HQ} act on two-dimensional spaces recording the outcomes. The traces in Eq.~\eqref{eq:HQmoments} instead use the physical time legs of $H^{\mathrm L}$. The relation between the two kernels can also be written directly. On each physical leg, let $\Pi_q$ project onto the eigenspace of the measured observable $Q$ with eigenvalue
$q$. With $\Pi_{\mathbf q}=\Pi_{q_1}\otimes\Pi_{q_2}\otimes\Pi_{q_3}$, we have
\begin{equation}
H_Q=
\sum_{\mathbf q}
\tr(H^{\mathrm L}\Pi_{\mathbf q})
\,|\mathbf q\rangle\langle\mathbf q|.
\label{eq:HQfromHL}
\end{equation}
Thus $H_Q$ keeps all the moments of $Q$, but discards information about other observables. This restriction cannot increase the trace norm or the negativity. Since $H_Q$ reproduces the pair correlations, it cannot have less negativity than the optimal pair kernel.

More generally, suppose additional operators are measured for the same preparation and L\"uders protocol. Let $\mathcal A$ contain these operators and the three pair products, and let $\Nmin(\mathcal A)$ be the least negativity among Hermitian, unit-trace kernels reproducing all their averages. Then for $\mathcal A\subseteq\mathcal B$,
\begin{equation}
\begin{aligned}
\N(H^{\mathrm L})
&\geq\N(H_Q)\geq\N(\Hpair)=\Nmin,\\
\Nmin(\mathcal A)&\leq\Nmin(\mathcal B)\leq\N(H^{\mathrm L}).
\end{aligned}
\label{eq:hierarchy}
\end{equation}
The first line compares three descriptions of the same L\"uders process. The second follows because adding measured averages cannot lower the minimum, while $H^{\mathrm L}$ itself reproduces every included average. If the measurements determine the full kernel, the upper value is attained. Supplementary Sections~S3 and S8 give the proofs.

For sequential L\"uders measurements, every three-time LG combination obeys the known upper bound of three halves, independently of the system dimension and Hamiltonian ~\cite{Budroni:2013pqf,PhysRevLett.113.050401}. Combining this bound with Eq.~\eqref{eq:main} gives
\begin{equation}
K_\alpha^{\mathrm L}\leq\frac32,
\qquad
\Nmin^{\mathrm L}\leq\frac18.
\label{eq:luders}
\end{equation}
This ceiling concerns the negativity required by the LG pair correlations, not the total negativity of a fuller kernel. Supplementary Section~S4 gives the short vector proof.
\begin{figure}[t]
\centering
\includegraphics[width=\columnwidth]{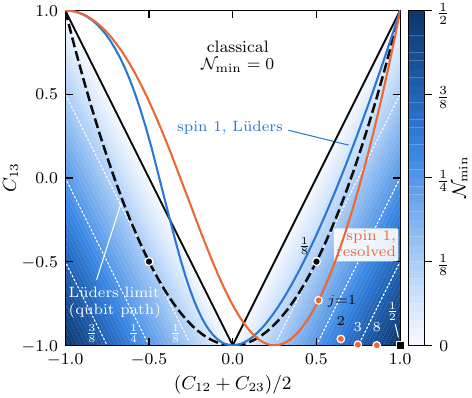}
\caption{Minimum negativity in the symmetric correlation plane. The colour map gives $\Nmin$ for $C_{12}=C_{23}=x$ versus $C_{13}$. The black V bounds the classical region in this plane; white dashed lines mark one eighth, one quarter and three eighths. The dashed parabola is the lower L\"uders boundary, traced by the equally spaced precessing qubit. Its black dots mark the L\"uders maximum of one eighth. The blue and orange curves show spin-one L\"uders and resolved protocols with equal intervals. These curves and the higher-spin maxima are plotted at their average pair correlation $x=(C_{12}+C_{23})/2$. Their minimum negativity equals that of the symmetric projection because $K_2$ and $K_3$ are not violated. Orange dots give resolved maxima for spins one, two, three and eight, approaching the corner with minimum negativity one half (square). Supplementary Section~S4 explains the projection.}
\label{fig:map}
\end{figure}

\begin{figure}[t]
\centering
\includegraphics[width=\columnwidth]{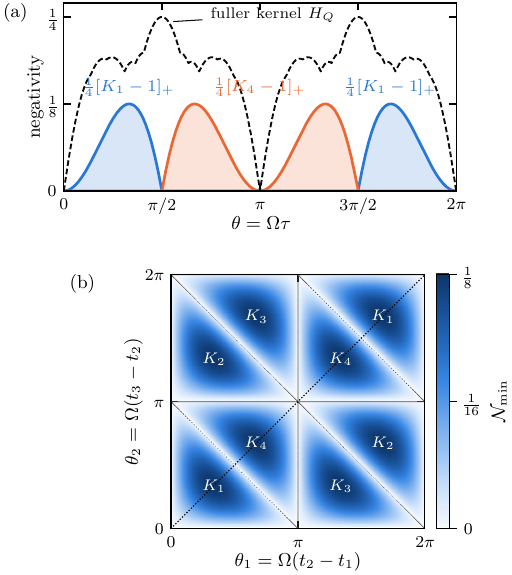}
\caption{Different LG inequalities reveal the qubit negativity in different time ranges. (a) Equal intervals with precession angle $\theta=\Omega\tau$. Shading shows the minimum negativity, revealed by $K_1$ when the cosine is positive (blue) and by $K_4$ when it is negative (orange). The dashed curve is the negativity of $H_Q$ for the state $(\id+\sigma_y)/2$ at the first time. It retains additional moments and never falls below the minimum. (b) Unequal intervals, with $\theta_1=\Omega(t_2-t_1)$ and $\theta_2=\Omega(t_3-t_2)$. Each of the eight regions is associated with one violated inequality and peaks at one eighth. The dotted diagonal is the equal-interval line of panel (a). The formulas for both panels are given in Supplementary Section~S5.}
\label{fig:qubit}
\end{figure}

\runin{Negativity from qubit measurements}
Consider a spin half evolving under $\Omega\sigma_x/2$, with $Q=\sigma_z$. The anticommutator of this observable at two times is proportional to the identity. Sequential L\"uders pair correlations are therefore independent of the initial state. For equal intervals, define the precession angle $\theta=\Omega\tau$ and write $c=\cos\theta$. Then
\begin{equation}
\begin{aligned}
C_{12}=C_{23}&=c,\qquad C_{13}=2c^2-1,\\
\Nmin&=\frac12|c|(1-|c|).
\end{aligned}
\label{eq:qubitN}
\end{equation}
Whenever the minimum is nonzero, $K_1$ is violated when the cosine is positive and $K_4$ when it is negative (Fig.~\ref{fig:qubit}(a)). A plot of $K_1$ alone would miss the latter intervals. The other two inequalities are never violated for equal intervals. The largest minimum negativity is one eighth, reached when the cosine is plus or minus one half. Unequal intervals allow all four inequalities to be violated in different regions. Figure~\ref{fig:qubit}(b) contains eight such regions, each reaching the same bound.

For this qubit, the pair correlations are independent of the initial state, but the additional moments retained by $H_Q$ need not be. Prepare the state $(\id+\sigma_y)/2$ at the first measurement time and choose equal intervals with $\theta=\Omega\tau=\pi/2$. All four LG inequalities then hold, and $\Hpair$ is positive. The additional moments are $m_1=m_3=0$, $m_2=1$ and $T=0$. Including them in Eq.~\eqref{eq:HQ} gives two eigenvalues of $-1/8$, so $\N(H_Q)=1/4$ (Fig.~\ref{fig:qubit}(a), dashed). This negativity is required to reproduce the additional moments, not an LG violation. By contrast, the full kernel $H^{\mathrm L}$ changes by unitary conjugation, so its eigenvalues and total negativity remain unchanged. Supplementary Sections~S2, S3 and S5 give the calculations.

The LG test uses only pair correlations of $Q=\sigma_z$. More information about correlations across time can be obtained by measuring another observable that anticommutes with $Q$, such as $S=\sigma_x$. These observables correspond to incompatible measurements. In separate runs with the same preparation, we measure $S$ only at the middle time, or measure $Q,S,Q$ at the three successive times. The two additional averages are included alongside the LG pair correlations in an extended kernel.

Incompatible measurements have previously been used to witness temporal negativity~\cite{Liu:2025dzl}. Here we determine the exact minimum over all Hermitian, unit-trace kernels on the three qubit time legs reproducing the five specified averages, and construct a kernel attaining it. For the initial state $(\id+\sigma_x)/2$ and $\theta=\pi/3$, this minimum is approximately 0.229, compared with 0.125 from the LG correlations alone. The LG violation is unchanged. Supplementary Section~S8 gives the closed-form result, the attaining kernel and the additional inequalities.

\runin{Beyond the L\"uders LG bound} A finer measurement can distinguish individual levels within an outcome subspace while reporting only the same two values. This von Neumann-type update~\cite{vonNeumann1932} disturbs the state differently and can exceed the L\"uders bound~\cite{PhysRevLett.113.050401}. Equation~\eqref{eq:main} still applies because it uses only the reported pair correlations.

Consider a spin $j$ precessing under $\Omega J_x$, initially prepared in $|j,-j\rangle$. The first measurement is made at $t_1=\pi/\Omega$, when the state is $|j,+j\rangle$ up to a phase. The subsequent measurement intervals are equal, with $\theta=\Omega\tau$. Only the lowest magnetic level reports minus one; every other level reports plus one. The L\"uders measurement projects onto these two outcome subspaces, whereas the resolved measurement projects onto each magnetic level separately.

For spin one and equal intervals, both protocols give the same $C_{12}$ and $C_{13}$, but the resolved value of $C_{23}$ is larger by $\tfrac12\sin^4\theta$. At the second time, the state contains the zero and positive magnetic levels, both reporting plus one. The L\"uders measurement preserves their relative phase; the resolved measurement removes it and changes the subsequent correlation. The Hamiltonian is the same in both cases. The violation increases because the measurements leave different states for the subsequent evolution. Supplementary Section~S6 gives the correlation formulas and the density matrices showing which coherence survives.

The resolved spin-one maximum certifies negativity about 0.189, exceeding the universal L\"uders bound. Unequal intervals raise it to about 0.227. Table~\ref{tab:spin} shows the equal-interval results for higher spins. For quarter-turn intervals, the first two pair correlations approach plus one and the first-to-third correlation is minus one. The LG combination consequently approaches its algebraic maximum of three, and the certified negativity approaches one half (Fig.~\ref{fig:map}). Supplementary Section~S6 proves this limit. These values describe the stated spin family, not global optima over all protocols of a given dimension.

\begin{table}[t]
\caption{Measurement resolution increases the certified negativity. Largest $K_1$ in the stated spin family for equal intervals, maximized over the precession angle, with L\"uders (L) or resolved (R) measurements. The final column is the minimum negativity certified by the resolved violation. The limiting row follows from the quarter-turn construction in Supplementary Section~S6.}
\label{tab:spin}
\begin{ruledtabular}
\begin{tabular}{cccc}
$j$ & $K_{1,\max}^{\mathrm L}$ & $K_{1,\max}^{\mathrm R}$ & $\Nmin^{\mathrm R}$\\
\hline
$1/2$ & 1.5000 & 1.5000 & 0.1250\\
$1$   & 1.3481 & 1.7565 & 0.1891\\
$3/2$ & 1.2053 & 2.0367 & 0.2592\\
$2$   & 1.1140 & 2.2532 & 0.3133\\
$3$   & 1.0325 & 2.4908 & 0.3727\\
$5$   & 1.0023 & 2.6437 & 0.4109\\
$8$   & 1.0000 & 2.7200 & 0.4300\\
$\to\infty$ & --- & $\to3$ & $\to1/2$
\end{tabular}
\end{ruledtabular}
\end{table}

\runin{LG tests at multiple measurement times}
The exact minimum extends to a cycle of any number of measurement times. The measured pairs connect successive times and the first and last times. Each LG combination is a signed sum of these pair correlations, with an odd number of minus signs~\cite{Mawby:2021mdc,Budroni:2013pqf}. The witness argument gives lower and upper bounds for every compatible kernel, while an explicit construction of history weights gives,
\begin{equation}
\begin{gathered}
n-2-\Delta_n[1+\N(H)]
\leq K_\sigma
\leq n-2+\Delta_n\N(H),\\
\begin{aligned}
\Nmin^{(n)}
&=\frac{\big[\max_\sigma K_\sigma-(n-2)\big]_+}{\Delta_n},\\
\Delta_n&=
\begin{cases}
2(n-2),&n\ \text{even},\\
2(n-1),&n\ \text{odd}.
\end{cases}
\end{aligned}
\end{gathered}
\label{eq:mainn}
\end{equation}
Here $\sigma$ labels the allowed sign patterns. For even $n$, the lower bound is equivalent to the upper bound for the opposite sign pattern and is already included in the maximization.

For every $n\geq3$, the least compatible negativity is nonzero if and only if at least one cycle LG inequality is violated. Both the bound and its attainment depend only on the measured cycle correlations, so the relation holds for any system dimension, initial state, dynamics or measurement rule.

The minimum follows from an explicit construction. For a violated inequality, positive weights are placed only on histories where its witness is largest, and negative weights only where it is smallest. Their total negative weight attains Eq.~\eqref{eq:mainn}. If no inequality is violated, positive weights suffice. For four or more times, setting every unmeasured moment to zero no longer generally gives the minimum; those moments must be chosen appropriately. Supplementary Section~S7 gives both proofs and the minimizing weights. As an example, an equally spaced qubit with eighth-turn intervals reaches the four-time L\"uders maximum of $2\sqrt2$~\cite{Budroni:2013pqf}, certifying minimum negativity about 0.207.

\runin{Discussion}
Our bounds connect LG violations to the spectrum of every Hermitian, unit-trace spacetime kernel reproducing the measured correlations. For any number of measurement times, the least compatible negativity is nonzero if and only if at least one of the corresponding LG inequalities is violated. The explicit kernels attaining this minimum identify how much negativity the measured correlations require, rather than the negativity introduced by an arbitrary choice of unmeasured moments.

Under unitary evolution, this minimum negativity can track changes in the measured LGI violation even while the full L\"uders kernel retains a fixed spectrum. Comparing L\"uders and resolved measurements of the same precessing spin shows how the measurement update affects the negativity required by the LG correlations. Resolving individual levels changes the subsequent correlations and can increase the negativity they require, even while removing coherence within an outcome subspace.

The result for incompatible qubit observables suggests a way to investigate temporal negativity without increasing an LG violation. Additional measurement directions can increase the lower bound on negativity even when the original LG correlations remain unchanged. A concrete next problem is to determine the smallest set of experimentally accessible temporal averages that certifies the full kernel's negativity, or a specified fraction of it, for a fixed process. For interacting systems, the same question can be asked using measurements confined to small spatial subsystems. Their kernels can have time-dependent spectra even when the complete-system spectrum is fixed. Establishing tight bounds from local temporal measurements would provide a direct connection between experimentally accessible correlations and the subsystem negativity studied in timelike entanglement.

\paragraph{Acknowledgments---} The authors would like to thank Taishi Kawamoto, Alexey Milekhin, Krishanu Roychowdhury, Ashish Shukla and Tadashi Takayanagi for useful discussions and comments. A.K. acknowledges the support of the Humboldt Research Fellowship for Experienced Researchers by the Alexander von Humboldt Foundation and the hospitality of Theoretical Physics III, Department of Physics and Astronomy, Julius-Maximilians-Universit\"{a}t W\"{u}rzburg, and the support from the ICTP through the Associates Programme (2024--2030) during the course of this work. R.N.D.'s work leading to this publication was supported by the PRIME programme of the German Academic Exchange Service (DAAD) with funds from the German Federal Ministry of Research, Technology and Space (BMFTR). R.N.D. is also supported by Germany's Excellence Strategy through the W\"urzburg-Dresden Cluster of Excellence on Complexity, Topology and Dynamics in Quantum Matter---ctd.qmat (EXC 2147, project-id 390858490).

\clearpage
\onecolumngrid

\setcounter{equation}{0}
\setcounter{section}{0}
\setcounter{table}{0}
\setcounter{figure}{0}

\renewcommand{\theequation}{S\arabic{equation}}
\renewcommand{\thesection}{S\arabic{section}}
\renewcommand{\thetable}{S\arabic{table}}
\renewcommand{\thefigure}{S\arabic{figure}}

\renewcommand{\theHequation}{supp.\arabic{equation}}
\renewcommand{\theHsection}{supp.\arabic{section}}
\renewcommand{\theHtable}{supp.\arabic{table}}
\renewcommand{\theHfigure}{supp.\arabic{figure}}

\section*{Supplementary Information for
``Exact operator bounds on Leggett--Garg violations''}

Equations, figures and tables numbered with an S prefix belong to this Supplementary Information. References to figures and tables in the main text are stated explicitly. Throughout, $Q=\Pi_+-\Pi_-$ is the measured two-valued observable, $\Pi_\pm$ are its outcome projectors, and $[x]_+=\max(x,0)$.

\tableofcontents

\section{Measured correlations}
\label{sec:operational}

In an LG experiment each pair of times $(t_i,t_j)$, $i<j$, is measured in its own runs. Let $\rho_i=U(t_i)\rho_0U(t_i)^\dagger$ be the state just before $t_i$, and $U_{ji}=U(t_j-t_i)$. For an instrument that, at $t_i$, applies the orthogonal projectors $\{P_a\}$ and reports the value $q_a\in\{\pm1\}$, the pair distribution and correlation are
\begin{equation}
\begin{aligned}
p_{ij}(q_i,q_j)
&=\sum_{a:\,q_a=q_i}\ \sum_{b:\,q_b=q_j}
\tr\!\left[P_bU_{ji}P_a\rho_iP_aU_{ji}^\dagger\right],\\
C_{ij}
&=\sum_{a,b}q_aq_b\,\tr\!\left[P_bU_{ji}P_a\rho_iP_aU_{ji}^\dagger\right].
\end{aligned}
\label{eq:Cop}
\end{equation}

\paragraph*{L\"uders update.} Here $\{P_a\}=\{\Pi_+,\Pi_-\}$. With $\Pi_\pm=(\id\pm Q)/2$,
\begin{equation}
\sum_{q=\pm1}q\,\Pi_q\rho\,\Pi_q=\Pi_+\rho\Pi_+-\Pi_-\rho\Pi_-=\tfrac12(Q\rho+\rho Q),
\label{eq:jordan}
\end{equation}
and therefore, in the Heisenberg picture $Q(t)=U(t)^\dagger QU(t)$,
\begin{equation}
C^{\mathrm L}_{ij}=\tfrac12\tr\!\left(\rho_0\{Q(t_i),Q(t_j)\}\right)=\mathrm{Re}\,\tr\!\left(\rho_0\,Q(t_i)Q(t_j)\right).
\label{eq:CL}
\end{equation}

\paragraph*{Resolved (von Neumann) update.} Here the $P_a$ are finer projectors, e.g., onto single levels, with $\sum_{a:\,q_a=\pm1}P_a=\Pi_\pm$. Equation~\eqref{eq:CL} then fails in general, because the first measurement also removes coherence \emph{within} each eigenspace of $Q$. None of our bounds uses Eq.~\eqref{eq:CL}. They use only the three numbers $C_{ij}$, however they were produced. In all cases $|C_{ij}|\le1$, since $C_{ij}$ is the average of a product of two numbers $\pm1$.

\section{The L\"uders spacetime kernel}
\label{sec:kernel}

\paragraph*{Construction.} Let $\{|k\rangle\}_{k=1}^d$ be a basis of the system and $E_{kl}=|k\rangle\langle l|$. Following the reconstruction procedure of Ref.~\cite{Das:2025fcd}, define the Jordan insertion $\mathcal J_A(X)=\frac12(AX+XA)$ and the evolution channel $\mathcal E_{j\leftarrow i}(X)=U_{ji}XU_{ji}^\dagger$. The nested (sequential) three-time functional
\begin{equation}
\begin{aligned}
F(O_1,O_2,O_3)
&=\tr\!\Big[O_3\,\mathcal E_{3\leftarrow2}
\Big(\mathcal J_{O_2}\big(\mathcal E_{2\leftarrow1}
(\mathcal J_{O_1}(\rho_1))\big)\Big)\Big]\\
&=\tfrac14\tr\!\left(\rho_0\{\{O_3(t_3),O_2(t_2)\},O_1(t_1)\}\right).
\end{aligned}
\label{eq:F}
\end{equation}
is linear in each argument. The kernel
\begin{equation}
H^{\mathrm L}=\sum_{k,l,m,n,r,s}F(E_{kl},E_{mn},E_{rs})\;E_{lk}\otimes E_{nm}\otimes E_{sr}
\label{eq:HL}
\end{equation}
satisfies $\tr[H^{\mathrm L}(O_1\otimes O_2\otimes O_3)]=F(O_1,O_2,O_3)$ for all $O_i$, because $\tr(E_{lk}E_{k'l'})=\delta_{kk'}\delta_{ll'}$.

\paragraph*{Properties.} (i) For Hermitian $O_i$ the right-hand side of Eq.~\eqref{eq:F} is real, since nested anticommutators of Hermitian operators are Hermitian. Products of Hermitian operators span the Hermitian operators on the three legs, so $H^{\mathrm L}$ is Hermitian. (ii) $\tr H^{\mathrm L}=F(\id,\id,\id)=1$. (iii) Because $\mathcal J_{\id}$ is the identity map, inserting $\id$ on a leg removes that measurement, and $\tr[H^{\mathrm L}(Q\otimes Q\otimes\id)]=C^{\mathrm L}_{12}$, and similarly for $C^{\mathrm L}_{23}$ and $C^{\mathrm L}_{13}$. (iv) By Eq.~\eqref{eq:jordan}, $F(Q,Q,Q)$ is the average of $q_1q_2q_3$ in a run with L\"uders measurements at all three times. For a qubit and Pauli observables, $F$ is the expectation value that defines the pseudo-density matrix of Ref.~\cite{Fitzsimons:2013gga}, so $H^{\mathrm L}$ coincides with it.

\paragraph*{The spectrum does not depend on the dynamics.} Write $O(t)=U(t)^\dagger OU(t)$ and let $H^{\mathrm L}_0$ be the kernel built from $\rho_0$ with no evolution, $\tr[H_0^{\mathrm L}(A\otimes B\otimes C)]=\frac14\tr(\rho_0\{\{C,B\},A\})$. Then
\begin{equation}
\begin{aligned}
\tr[H^{\mathrm L}(O_1\otimes O_2\otimes O_3)]
&=\tr\!\big[H^{\mathrm L}_0\,(O_1(t_1)\otimes O_2(t_2)\otimes O_3(t_3))\big],\\
H^{\mathrm L}&=\mathcal U\,H^{\mathrm L}_0\,\mathcal U^\dagger,\\
\mathcal U&=U(t_1)\otimes U(t_2)\otimes U(t_3).
\end{aligned}
\label{eq:covariance}
\end{equation}
A further unitary $V$ on $\rho_0$ acts as $V^{\otimes3}$ on $H^{\mathrm L}_0$. Hence the eigenvalues of $H^{\mathrm L}$, and in particular $\N(H^{\mathrm L})$, depend only on the eigenvalues of $\rho_0$. They are independent of the Hamiltonian and of all three measurement times. For a qubit with Bloch vector of length $r$ we find numerically that $\N(H^{\mathrm L})$ increases from $1$ at $r=0$ to its pure-state value at $r=1$, where the eight eigenvalues are
\begin{equation}
\begin{aligned}
\operatorname{spec}(H^{\mathrm L})
&=\Big\{-\tfrac12,\ -\tfrac{1}{2\sqrt2},\ -\tfrac{\sqrt3-1}{4},\
0,\ 0,\ \tfrac{1}{2\sqrt2},\ \tfrac{\sqrt3+1}{4},\ 1\Big\},\\
\N(H^{\mathrm L})&=\frac{1+\sqrt2+\sqrt3}{4}\simeq1.0366 .
\end{aligned}
\label{eq:HLspec}
\end{equation}
For $r=0$ the eigenvalues are $-\tfrac14$ and $\tfrac12$, four times each. This invariance holds whenever each leg is the complete system. It fails when legs are restricted to subsystems, which is the setting of timelike-entanglement studies~\cite{Milekhin:2025ycm,Guo:2025dtq,Das:2026ifj}, or when the kernel is read through the measured observable, as in Sec.~\ref{sec:three}.

\section{The three-time minimum}
\label{sec:three}

\paragraph*{Compatible kernels.} A kernel $H$ on legs $\mathcal H_1\otimes\mathcal H_2\otimes\mathcal H_3$ (any dimensions) with two-valued leg observables $Q_i$, meaning $Q$ on leg $i$ and the identity on the other legs, $Q_i^2=\id$, is compatible with data $(C_{12},C_{23},C_{13})$ if $H=H^\dagger$, $\tr H=1$, and $\tr(HQ_iQ_j)=C_{ij}$. Its negativity is $\N(H)=\sum_{\lambda<0}|\lambda|=\tr H_-$, where $H=H_+-H_-$ with $H_\pm\ge0$ and $H_+H_-=0$. Since $\tr H_+-\tr H_-=1$ and $\|H\|_1=\tr H_++\tr H_-$, we have $\N(H)=(\|H\|_1-1)/2$.

\paragraph*{Witnesses.} The four witnesses are
\begin{equation}
\begin{aligned}
W_1&=Q_1Q_2+Q_2Q_3-Q_1Q_3, &\qquad W_2&=Q_1Q_2-Q_2Q_3+Q_1Q_3,\\
W_3&=-Q_1Q_2+Q_2Q_3+Q_1Q_3, &\qquad W_4&=-Q_1Q_2-Q_2Q_3-Q_1Q_3 .
\end{aligned}
\label{eq:Ws}
\end{equation}
The $Q_i$ commute, and on a joint eigenstate with eigenvalues $(q_1,q_2,q_3)$ we have, with $a=q_1q_2$ and $b=q_2q_3$, $W_1=a+b-ab\in\{1,-3\}$. The substitution $q_2\to-q_2$ maps $W_1\to W_4$, $q_3\to-q_3$ maps $W_1\to W_2$, and $q_1\to-q_1$ maps $W_1\to W_3$, so all four witnesses satisfy $-3\,\id\le W_\alpha\le\id$.

For normalized history weights, the first combination obeys
\begin{equation}
K_1=1-4\big[P(+,-,+)+P(-,+,-)\big].
\label{eq:zigzag}
\end{equation}
The two displayed histories give the witness value minus three, and all other histories give one. Their combined weight is therefore negative whenever the first LG inequality is violated. The same argument applies to the other sign patterns.

\paragraph*{Lower bound.} For any compatible $H$,
\begin{equation}
\begin{aligned}
K_\alpha
&=\tr(HW_\alpha)=\tr(H_+W_\alpha)-\tr(H_-W_\alpha)\\
&\leq\tr H_++3\tr H_-=1+4\N(H).
\end{aligned}
\label{eq:lower}
\end{equation}

\paragraph*{The pair kernel for arbitrary systems.}
The construction uses the measurement outcomes rather than
the full state space of the physical system. For each time,
introduce two orthonormal outcome states and define
\begin{equation}
Q|q\rangle=q|q\rangle,
\qquad q=\pm1.
\end{equation}
The physical system need not be a qubit. Its state, dynamics
and measurement rule determine the pair correlations, but
do not enter the following construction.

On the three outcome spaces, define
\begin{align}
\Hpair=\frac18\Big[&\id^{\otimes3}
+C_{12}(Q\otimes Q\otimes\id)
\nonumber\\
&+C_{23}(\id\otimes Q\otimes Q)
+C_{13}(Q\otimes\id\otimes Q)\Big].
\label{eq:pairconstruction}
\end{align}
The outcome observable is traceless and squares to the
identity. Orthogonality of the tensor-product operators
therefore gives
\begin{align}
\tr\Hpair&=1,
\\
\tr(\Hpair Q_iQ_j)&=C_{ij},
\qquad i<j.
\end{align}
Its single-time averages and three-time correlation vanish.
This is a choice of the moments left unspecified by the
LG test, not an assumption about the physical initial state.

The history states
\begin{equation}
|\mathbf q\rangle
=|q_1\rangle\otimes|q_2\rangle\otimes|q_3\rangle
\end{equation}
diagonalize the kernel. Their eigenvalues are
\begin{equation}
\lambda(\mathbf q)
=\frac18\left(
1+q_1q_2C_{12}+q_2q_3C_{23}+q_1q_3C_{13}
\right).
\label{eq:pairhistory}
\end{equation}
The pair products satisfy
\begin{equation}
(q_1q_2)(q_2q_3)(q_1q_3)=1,
\end{equation}
so they contain either zero or two minus signs.
Substituting the four possibilities gives
\begin{align}
\lambda(+,-,+)
&=\frac{1-C_{12}-C_{23}+C_{13}}8
=\frac{1-K_1}8,
\\
\lambda(+,-,-)
&=\frac{1-C_{12}+C_{23}-C_{13}}8
=\frac{1-K_2}8,
\\
\lambda(+,+,-)
&=\frac{1+C_{12}-C_{23}-C_{13}}8
=\frac{1-K_3}8,
\\
\lambda(+,+,+)
&=\frac{1+C_{12}+C_{23}+C_{13}}8
=\frac{1-K_4}8.
\end{align}
Reversing all three outcomes leaves every pair product
unchanged. Each eigenvalue consequently occurs twice,
\begin{equation}
\lambda_\alpha=\frac{1-K_\alpha}{8},
\qquad \alpha=1,\ldots,4,
\quad\text{each twice}.
\label{eq:pairspectrum}
\end{equation}

The kernel is positive precisely when all four LG
inequalities hold. Its total negative eigenvalue weight is
\begin{equation}
\N(\Hpair)=\frac14
\sum_{\alpha=1}^{4}[K_\alpha-1]_+.
\end{equation}
At most one combination can exceed its classical bound,
because any two distinct combinations satisfy
\begin{equation}
K_\alpha+K_\beta=\pm2C_{ij}\leq2,
\qquad \alpha\neq\beta.
\end{equation}
Therefore
\begin{equation}
\N(\Hpair)
=\frac14[\max_\alpha K_\alpha-1]_+.
\label{eq:Npair}
\end{equation}
The universal lower bound in Eq.~\eqref{eq:lower} applies
to every compatible kernel. Since the pair kernel attains
the resulting minimum,
\begin{equation}
\Nmin=\N(\Hpair)
=\frac14[\max_\alpha K_\alpha-1]_+.
\end{equation}
This proves the exact three-time minimum stated in the main text. In particular,
\begin{equation}
\N(\Hpair)>0
\quad\Longleftrightarrow\quad
K_\alpha>1\ \text{for some }\alpha.
\end{equation}

The construction also establishes attainment on the full
physical time legs. Choose one normalized physical state
in each outcome subspace and let the isometry $V$ map
the outcome states to these states,
\begin{equation}
V|q\rangle=|\phi_q\rangle,
\qquad
\Pi_q|\phi_q\rangle=|\phi_q\rangle.
\end{equation}
The embedded kernel
\begin{equation}
\widetilde H_{\mathrm{pair}}
=V^{\otimes3}\Hpair(V^\dagger)^{\otimes3}
\end{equation}
has the same pair correlations and nonzero eigenvalues.
Any additional eigenvalues are zero. The minimum is
therefore attained for physical systems of arbitrary
dimension and arbitrary outcome degeneracies.

The minimum value is fixed by the pair correlations,
but the operator attaining it need not be unique.
Changing the physical preparation can change the
correlations and hence the numerical minimum.
What remains independent of the preparation is the
exact relation between them.

\paragraph*{Pair kernel and minimal negativity.} Let $H$ be compatible, with leg projectors $\Pi^{(i)}_{q}$. Define the measurement map onto outcome registers,
\begin{equation}
\Phi(H)=\sum_{q_1,q_2,q_3}\tr\!\Big[H\,\big(\Pi^{(1)}_{q_1}\otimes\Pi^{(2)}_{q_2}\otimes\Pi^{(3)}_{q_3}\big)\Big]\,|q_1q_2q_3\rangle\langle q_1q_2q_3| .
\label{eq:Phi}
\end{equation}
$\Phi$ is linear, positive, and trace preserving, and it preserves the $Q_iQ_j$ correlations on the outcome registers. For such a map, the trace norm cannot increase on a Hermitian operator. Writing $X=X_+-X_-$ gives $\|\Phi(X)\|_1\le\tr\Phi(X_+)+\tr\Phi(X_-)=\|X\|_1$. So $\N(\Phi(H))\le\N(H)$. The diagonal operator $\Phi(H)$ has the moment expansion
\begin{equation}
\begin{aligned}
\Phi(H)=\tfrac18\Big[&\id^{\otimes3}+m_1(Q\otimes\id\otimes\id)
+m_2(\id\otimes Q\otimes\id)+m_3(\id\otimes\id\otimes Q)\\
&+C_{12}(Q\otimes Q\otimes\id)+C_{23}(\id\otimes Q\otimes Q)\\
&+C_{13}(Q\otimes\id\otimes Q)+T(Q\otimes Q\otimes Q)\Big],
\end{aligned}
\label{eq:moments}
\end{equation}
with $m_i=\tr(HQ_i)$ and $T=\tr(HQ_1Q_2Q_3)$. Let $X_{\mathrm{flip}}|q_1q_2q_3\rangle=|-q_1,-q_2,-q_3\rangle$. The average $\frac12[\Phi(H)+X_{\mathrm{flip}}\Phi(H)X_{\mathrm{flip}}]$ removes the odd terms and equals $\Hpair$; by convexity of the trace norm it has no more negativity than $\Phi(H)$. Hence
\begin{equation}
\N(H)\ \ge\ \N\big(\Phi(H)\big)\ \ge\ \N(\Hpair)=\Nmin
\label{eq:chain}
\end{equation}
for every compatible $H$, of any dimension and whatever the ranks of the $\Pi^{(i)}_q$. For the L\"uders kernel, $\Phi(H^{\mathrm L})$ is the outcome-level kernel $H_Q$ discussed in the main text. Its single-time averages and sequential three-time correlation are
\begin{equation}
m_i=\tr(H^{\mathrm L}Q_i)=\langle Q(t_i)\rangle,
\qquad
T=\tr(H^{\mathrm L}Q_1Q_2Q_3)=F(Q,Q,Q).
\end{equation}
The map preserves these moments, so they are also obtained by tracing $H_Q$ with the corresponding outcome observables.

\paragraph*{Additional moments and excess negativity.}
The fuller outcome kernel also retains the single-time
averages and three-time correlation. Its history weights
are
\begin{equation}
w(\mathbf q)=\lambda(\mathbf q)
+\frac18\left(
m_1q_1+m_2q_2+m_3q_3+Tq_1q_2q_3
\right),
\end{equation}
where the pair-kernel eigenvalue is given in
Eq.~\eqref{eq:pairhistory}.

Reversing all three outcomes preserves the pair terms
but reverses the additional terms. The two weights
associated with each LG combination can therefore
be written as
\begin{equation}
w_{\alpha,\pm}
=\frac{1-K_\alpha\pm d_\alpha}{8},
\label{eq:splitweights}
\end{equation}
with
\begin{equation}
\begin{aligned}
d_1&=m_1-m_2+m_3-T,&
d_2&=m_1-m_2-m_3+T,\\
d_3&=m_1+m_2-m_3-T,&
d_4&=m_1+m_2+m_3+T.
\end{aligned}
\end{equation}
Their sum is fixed by the pair correlations,
\begin{equation}
w_{\alpha,+}+w_{\alpha,-}
=\frac{1-K_\alpha}{4},
\end{equation}
but their individual signs also depend on the additional
moments. A nonnegative sum does not prevent one of the
two weights from being negative.

The contribution of each history pair to the negativity
is
\begin{align}
\nu_\alpha
&=[-w_{\alpha,+}]_++[-w_{\alpha,-}]_+
\nonumber\\
&=\frac18\left[
\max\bigl\{|1-K_\alpha|,|d_\alpha|\bigr\}
-(1-K_\alpha)\right].
\end{align}
Subtracting the corresponding pair-kernel contribution
and summing over the four pairs gives the exact excess,
\begin{equation}
\N(H_Q)-\Nmin
=\frac18\sum_{\alpha=1}^{4}
\bigl[|d_\alpha|-|1-K_\alpha|\bigr]_+.
\label{eq:excessnegativity}
\end{equation}
Thus the additional moments can increase the negativity
without changing any LG combination. They do not
necessarily increase it. The fuller outcome kernel
remains minimal precisely when
\begin{equation}
\N(H_Q)=\Nmin
\quad\Longleftrightarrow\quad
|d_\alpha|\leq|1-K_\alpha|
\ \text{for every }\alpha.
\label{eq:HQminimal}
\end{equation}
This explains both possibilities illustrated by the
qubit preparations in Sec.~\ref{sec:qubit}.

On a $d$-level physical leg the operator $d^{-3}[\id^{\otimes3}+C_{12}(Q\otimes Q\otimes\id)+C_{23}(\id\otimes Q\otimes Q)+C_{13}(Q\otimes\id\otimes Q)]$ is \emph{not} in general compatible with the data. If $\tr Q\ne0$, its trace is not one and its pair moments are not $C_{ij}$ (for $Q=\mathrm{diag}(-1,1,1)$ its trace is $1+(C_{12}+C_{23}+C_{13})/9$). Working on two-dimensional outcome registers, or with the map~\eqref{eq:Phi}, avoids this problem, and it is why no assumption such as $\tr Q=0$ is needed.

\paragraph*{Distance to classical descriptions.} For Hermitian $H$ with $\tr H=1$,
\begin{equation}
\min_{\sigma\ge0,\ \tr\sigma=1}\tfrac12\|H-\sigma\|_1=\N(H).
\label{eq:distance}
\end{equation}
Choosing $\sigma=H_+/(1+\N)$ gives the upper bound $\frac12\|H-\sigma\|_1=\frac12\big[\N+\N\big]=\N$. For the lower bound, let $P$ project onto the support of $H_-$. Since $\tr(H-\sigma)=0$, $\frac12\|H-\sigma\|_1\ge\tr[P(\sigma-H)]\ge\tr(PH_-)=\N$. Thus $\Nmin$ is the smallest trace distance between any compatible description and an ordinary density matrix; see also Refs.~\cite{Liu:2025dzl,Das:2026ifj}.

\paragraph*{Sampling cost.} Suppose the three correlations are simulated classically by drawing histories $q$ from $|P(q)|/\|P\|_1$ and recording $\mathrm{sign}[P(q)]\,\|P\|_1\,q_iq_j$, where $P$ are signed history weights reproducing the data and $\|P\|_1=\sum_q|P(q)|=1+2\N$. Each recorded number lies in $[-\|P\|_1,\|P\|_1]$, so by Hoeffding's inequality $M\ge2\|P\|_1^2\ln(2/\delta)/\epsilon^2$ samples estimate each $C_{ij}$ to accuracy $\epsilon$ with confidence $1-\delta$~\cite{Pashayan2015}. The smallest possible prefactor is $(1+2\Nmin)^2=(1+[K-1]_+/2)^2$, with $K=\max_\alpha K_\alpha$. It ranges from $1$ (classical data) to $(5/4)^2$ at the L\"uders ceiling and to $4$ at the corners of the cube.

\section{Geometry and the L\"uders ceiling}
\label{sec:geometry}

\paragraph*{Classical region.} The four deterministic correlation triples $(C_{12},C_{23},C_{13})=(1,1,1)$, $(1,-1,-1)$, $(-1,1,-1)$, $(-1,-1,1)$ are the vertices of a tetrahedron whose faces are $K_\alpha=1$. A triple admits a nonnegative history distribution if and only if it lies in this tetrahedron~\cite{Fine1982,DezaLaurent1997,Halliwell:2019uzy}; this is the case $\Nmin=0$ of Eq.~\eqref{eq:Npair}. The level set $\Nmin=\nu$ is the boundary of the tetrahedron $K_\alpha\le1+4\nu$, i.e., the original one scaled by $1+4\nu$ about the origin, intersected with the cube $|C_{ij}|\le1$. At $\nu=1/2$ it reaches the cube corners $(1,1,-1)$, $(1,-1,1)$, $(-1,1,1)$, $(-1,-1,-1)$, where one $K_\alpha=3$.

\paragraph*{L\"uders region.} For L\"uders updates, purify $\rho_0$ to $|\psi\rangle$ (with $Q(t)$ acting trivially on the purifying system) and set $v_i=Q(t_i)|\psi\rangle$. Then $\|v_i\|=1$ and, by Eq.~\eqref{eq:CL}, $C_{ij}=\mathrm{Re}\langle v_i|v_j\rangle$, which is the real inner product of $v_i$ and $v_j$ viewed as real vectors. The matrix
\begin{equation}
G=\begin{pmatrix}1&C_{12}&C_{13}\\ C_{12}&1&C_{23}\\ C_{13}&C_{23}&1\end{pmatrix}
\label{eq:gram}
\end{equation}
is therefore positive semidefinite. Conversely, every such $G$ is realized by a qubit with $Q(t_i)=\bm n_i\cdot\bm\sigma$ for suitable unit vectors $\bm n_i$ and dynamics. The set of positive semidefinite $G$ is known as the elliptope. The squared norm
\begin{equation}
0\leq\|v_1-v_2+v_3\|^2
=3-2C_{12}-2C_{23}+2C_{13}
=3-2K_1
\label{eq:luders-vector}
\end{equation}
gives the L\"uders bound $K_1\leq3/2$~\cite{Budroni:2013pqf,PhysRevLett.113.050401}. Changing the vector signs gives the same bound for the other LG combinations. The proof uses unit norms and does not require a particular Hamiltonian. Therefore L\"uders data always obey $\Nmin\le1/8$, independently of dimension.

\paragraph*{The parabola in Fig.~1 of the main text.} On the plane $C_{12}=C_{23}=x$, $C_{13}=y$, $\det G=(1-y)(1+y-2x^2)$, so $G\ge0$ if and only if $y\ge2x^2-1$. For a general triple, the swap $C_{12}\leftrightarrow C_{23}$ (relabeling $t_1\leftrightarrow t_3$) maps the elliptope and the tetrahedron to themselves. By convexity, a point $(C_{12},C_{23},y)$ in either set implies that the symmetric point $(x,x,y)$ with $x=(C_{12}+C_{23})/2$ is in the same set. The projection used in Fig.~1 of the main text thus shows the true L\"uders region $y\ge2x^2-1$ and the true classical region $y\ge2|x|-1$. Along it $K_1=2x-y$ and $K_4=-2x-y$ are exact, while $K_{2,3}=\pm(C_{12}-C_{23})+y$ are not displayed; for all curves shown in the figure, $K_{2,3}\le1$ wherever a violation occurs.

\paragraph*{More times.} For $n$ times the L\"uders correlations again form a positive semidefinite $n\times n$ matrix with unit diagonal. Maximizing the chain combination over such matrices gives $n\cos(\pi/n)$~\cite{Budroni:2013pqf}; we confirmed this value by semidefinite programming for $n=3,4,5,6,8$. With Eq.~\eqref{eq:minimum-n} the L\"uders ceiling on the $n$-time certified negativity is $[n\cos(\pi/n)-(n-2)]/\Delta_n$, i.e., $0.1250$, $0.2071$, $0.1306$, $0.1495$, $0.1089$, $0.1159$ for $n=3,\dots,8$.

\section{Precessing qubit}
\label{sec:qubit}

With Hamiltonian $\Omega\sigma_x/2$ and $Q=\sigma_z$, $Q(t)=\sigma_z\cos\Omega t+\sigma_y\sin\Omega t$ and $\{Q(t),Q(t')\}=2\cos[\Omega(t'-t)]\,\id$. Hence $C_{ij}=\cos[\Omega(t_j-t_i)]$ for every initial state. With $\theta_1=\Omega(t_2-t_1)$ and $\theta_2=\Omega(t_3-t_2)$,
\begin{equation}
K_1=\cos\theta_1+\cos\theta_2-\cos(\theta_1+\theta_2),\quad
K_2=\cos\theta_1-\cos\theta_2+\cos(\theta_1+\theta_2),
\end{equation}
\begin{equation}
K_3=-\cos\theta_1+\cos\theta_2+\cos(\theta_1+\theta_2),\quad
K_4=-\cos\theta_1-\cos\theta_2-\cos(\theta_1+\theta_2).
\end{equation}
Each $K_\alpha$ reaches $3/2$ at one point in each of two lobes of the torus $[0,2\pi)^2$. The maxima occur for $K_1$ at $(\theta_1,\theta_2)=(\pi/3,\pi/3)$ and $(5\pi/3,5\pi/3)$; $K_4$ at $(2\pi/3,2\pi/3)$ and $(4\pi/3,4\pi/3)$; $K_2$ at $(\pi/3,4\pi/3)$ and $(5\pi/3,2\pi/3)$; $K_3$ at $(2\pi/3,5\pi/3)$ and $(4\pi/3,\pi/3)$. The lobes are separated by the lines $\theta_1\in\{0,\pi\}$, $\theta_2\in\{0,\pi\}$ and $\theta_1+\theta_2\in\{\pi,2\pi,3\pi\}$, on which $\Nmin=0$ [Fig.~2b of the main text].

For equal intervals, write $c=\cos\theta$. The combinations and minimum are
\begin{equation}
\begin{aligned}
K_{1,4}&=1\pm2c(1\mp c),&
K_{2,3}&=\cos2\theta,\\
\Nmin&=\frac12|c|(1-|c|).
\end{aligned}
\label{eq:qubitminimum}
\end{equation}
The pair-kernel eigenvalues are $c(1+c)/4$ and $c(c-1)/4$, twice each, and $(1-c^2)/4$, four times. Thus the second and third inequalities cannot be violated for equal intervals. For unequal intervals they can be violated, as the formulas above show.

The single-time and three-time moments are instead
\begin{equation}
m_i=\tr[\rho_0Q(t_i)],
\qquad
T=\frac14\tr\!\left[\rho_0\{\{Q(t_3),Q(t_2)\},Q(t_1)\}\right]
=C_{23}m_1.
\label{eq:qubitodd}
\end{equation}
They can depend on the initial state, whereas the pair correlations do not in this qubit model. If the second-to-third interval is a quarter-turn, the three-time moment is zero for every initial state. The state dependence of the fuller outcome kernel then comes only from the single-time averages.

\paragraph*{Spin-up initial state.} Consider instead a qubit prepared in the positive eigenstate of the measured observable at the first measurement time. Then $m_1=1$, $m_2=c$, $m_3=\cos2\theta$, and $T=\tfrac14\langle\{\{Q(t_3),Q(t_2)\},Q(t_1)\}\rangle=c\,m_1=c$. All histories with $q_1=-1$ get zero weight, and for $q_1=+1$
\begin{equation}
P(+,q_2,q_3)=\tfrac14\left(1+c\,q_2+\cos2\theta\,q_3+c\,q_2q_3\right).
\end{equation}
The only possibly negative weights are $P(+,-,+)=c(c-1)/2$ and $P(+,+,+)=c(1+c)/2$, so $\N(H_Q)=|c|(1-|c|)/2=\Nmin$. For this preparation the outcome-level kernel is already minimal.

\paragraph*{Negativity without LG violation.} For $\rho(t_1)=(\id+\sigma_y)/2$ and $\theta=\pi/2$ one has $Q(t_1)=\sigma_z$, $Q(t_2)=\sigma_y$, $Q(t_3)=-\sigma_z$. Then $(m_1,m_2,m_3)=(0,1,0)$, $(C_{12},C_{23},C_{13})=(0,0,-1)$, and $T=0$ because $\{\sigma_z,\sigma_y\}=0$. So $H_Q=\tfrac18[\id^{\otimes3}+(\id\otimes Q\otimes\id)-(Q\otimes\id\otimes Q)]$ with eigenvalues $-\frac18$ (twice), $\frac18$ (four times), $\frac38$ (twice), and $\N(H_Q)=\frac14$, while $(K_1,K_2,K_3,K_4)=(1,-1,-1,1)$ and $\Nmin=0$. The negativity is forced by combining $m_2=1$, obtained without a measurement at $t_1$, with the other moments. A L\"uders measurement at $t_1$ dephases the state in the $\sigma_z$ basis and sets $\langle Q(t_2)\rangle$ to $0$. This is a violation of no-signaling in time~\cite{Kofler2013-nw,Halliwell_2016}, which the LG inequalities cannot detect~\cite{Clemente:2016jgc}. The dashed curve in Fig.~2a of the main text shows $\N(H_Q)$ for this state as a function of $\theta$. The full kernel has $\N(H^{\mathrm L})=(1+\sqrt2+\sqrt3)/4$ for all $\theta$, Eq.~\eqref{eq:HLspec}.

\section{Spin-\texorpdfstring{$j$}{j} measurement protocols}
\label{sec:spin}

The Hamiltonian is $\Omega J_x$, the initial state is $|j,-j\rangle$, and the first measurement is at $\Omega t_1=\pi$, where the state is $|j,+j\rangle$ up to a phase. The reported value is $q_m=-1$ for $m=-j$ and $+1$ otherwise. The L\"uders instrument uses $\Pi_-=|j,-j\rangle\langle j,-j|$ and $\Pi_+=\id-\Pi_-$; the resolved instrument uses the $2j+1$ projectors $|j,m\rangle\langle j,m|$. Because the state at $t_1$ is an eigenstate of every projector, both instruments give the same $C_{12}$ and $C_{13}$; they differ only in $C_{23}$, whose first measurement acts on a superposition.

\paragraph*{Spin 1.} With $c=\cos\theta$ and the transition probabilities $|\langle m|e^{-i\theta J_x}|m'\rangle|^2$, Eq.~\eqref{eq:Cop} gives
\begin{equation}
\begin{aligned}
C^{\mathrm R}_{12}&=c+\tfrac12\sin^2\theta,\\
C^{\mathrm R}_{23}&=c+\tfrac18\sin^22\theta,\\
C^{\mathrm R}_{13}&=\cos2\theta+\tfrac12\sin^22\theta,
\end{aligned}
\end{equation}
and $C^{\mathrm L}_{12}=C^{\mathrm R}_{12}$, $C^{\mathrm L}_{13}=C^{\mathrm R}_{13}$, $C^{\mathrm L}_{23}=C^{\mathrm R}_{23}-\frac12\sin^4\theta$. Thus
\begin{equation}
\begin{aligned}
K_1^{\mathrm R}&=\tfrac32+2c-4c^2+\tfrac32c^4,\\
K_1^{\mathrm L}&=1+2c-3c^2+c^4,\\
K_1^{\mathrm R}-K_1^{\mathrm L}
&=\tfrac12(1-c^2)^2=\tfrac12\sin^4\theta .
\end{aligned}
\end{equation}
Setting $dK_1^{\mathrm R}/dc=0$ gives $(c-1)(3c^2+3c-1)=0$, so the interior maximum is at $c=(\sqrt{21}-3)/6$, i.e., $\theta\simeq1.30388$, with $K_1^{\mathrm R}=1.756502$ and $\Nmin=0.189126$. For L\"uders, $(c-1)(2c^2+2c-1)=0$ gives $c=(\sqrt3-1)/2$ and $K_1^{\mathrm L}=1.348076$, below the L\"uders ceiling, as it must be. In the resolved run the relative phase between $|1,0\rangle$ and $|1,+1\rangle$ at $t_2$ is destroyed; the L\"uders run keeps it. This is the only difference, and it accounts for the $\frac12\sin^4\theta$.

\paragraph*{Coherence under the two measurement updates.}
For spin~$1$, both $|1,+1\rangle$ and $|1,0\rangle$ give the
reported outcome $q=+1$, whereas $|1,-1\rangle$ gives $q=-1$.
Thus
\begin{equation}
\Pi_+=|1,+1\rangle\langle1,+1|+|1,0\rangle\langle1,0|,
\qquad
\Pi_-=|1,-1\rangle\langle1,-1|.
\end{equation}
Immediately before the measurement at $t_2$, the state is
\begin{equation}
|\psi_2\rangle=e^{-i\theta J_x}|1,+1\rangle
=a|1,+1\rangle-is|1,0\rangle-b|1,-1\rangle,
\end{equation}
where
\begin{equation}
a=\frac{1+\cos\theta}{2},\qquad
b=\frac{1-\cos\theta}{2},\qquad
s=\frac{\sin\theta}{\sqrt2}.
\end{equation}
Write $\rho_2=|\psi_2\rangle\langle\psi_2|$. Coherence between
the spin states appears in the off-diagonal entries of this
density matrix.

The L\"uders projector acts on both states in the $+1$ subspace
together,
\begin{equation}
\Pi_+|\psi_2\rangle
=a|1,+1\rangle-is|1,0\rangle.
\end{equation}
Their relative phase is unchanged. Summing over the two
measurement outcomes gives, in the basis
$(|1,+1\rangle,|1,0\rangle,|1,-1\rangle)$,
\begin{equation}
\rho_{\mathrm L}
=\Pi_+\rho_2\Pi_++\Pi_-\rho_2\Pi_-
=
\begin{pmatrix}
a^2&ias&0\\
-ias&s^2&0\\
0&0&b^2
\end{pmatrix}.
\end{equation}
The entries $ias$ and $-ias$ survive because both states belong
to the same outcome subspace. Coherence between different
outcome subspaces is removed.

The resolved measurement instead projects onto each spin state
separately. With $P_m=|1,m\rangle\langle1,m|$, summing over its
outcomes gives
\begin{equation}
\rho_{\mathrm R}
=\sum_{m=-1}^{1}P_m\rho_2P_m
=
\begin{pmatrix}
a^2&0&0\\
0&s^2&0\\
0&0&b^2
\end{pmatrix}.
\end{equation}
It therefore removes the coherence between $|1,+1\rangle$
and $|1,0\rangle$ that the L\"uders measurement preserves.
For example, at $\theta=\pi/2$,
\begin{equation}
\langle1,+1|\rho_{\mathrm L}|1,0\rangle
=\frac{i}{2\sqrt2},
\qquad
\langle1,+1|\rho_{\mathrm R}|1,0\rangle=0.
\end{equation}

Both measurements give the same probabilities at $t_2$,
$p(+)=a^2+s^2$ and $p(-)=b^2$. They nevertheless leave different
states for the subsequent evolution. Reporting only $q=\pm1$
after resolving the individual spin states does not restore the
removed coherence.

\paragraph*{Higher spins.} We evaluated Eq.~\eqref{eq:Cop} with $J_\pm$ matrices and maximized over $\theta$ for equal intervals. All correlations are symmetric under $\theta\to2\pi-\theta$. For spins of one and higher, the resolved updates violate only $K_1$; the spin-half case can also violate $K_4$. With L\"uders updates $K_4$ is also violated in some intervals of $\theta$, always below $3/2$. The results are listed in Table~\ref{tab:spinS}. The resolved values approach the large-$j$ form $3-\sqrt{2/(\pi j)}$ of Ref.~\cite{PhysRevLett.113.050401}, which they match to within $0.003$ for $j=5$ and $j=8$. For comparison, the table also gives the maximum over unequal intervals $(\theta_1,\theta_2)$ in the same family, found by a grid search followed by local optimization.

\begin{table}[h]
\caption{Spin-$j$ family. Equal-interval columns give $K^{\mathrm L}$, $K^{\mathrm R}$, $\theta^\ast$ (position of the resolved maximum in $(0,\pi)$), and $\Nmin=(K^{\mathrm R}-1)/4$. The final two columns give the best resolved value and its minimum negativity for unequal intervals.}
\label{tab:spinS}
\begin{ruledtabular}
\begin{tabular}{ccccccc}
 & \multicolumn{4}{c}{Equal intervals} & \multicolumn{2}{c}{Unequal intervals}\\
$j$ & $K^{\mathrm L}_{\max}$ & $K^{\mathrm R}_{\max}$ & $\theta^\ast$ & $\Nmin$ & $K^{\mathrm R}_{\max}$ & $\Nmin$\\
\hline
$1/2$ & 1.50000 & 1.50000 & 1.0472 & 0.12500 & 1.50000 & 0.12500\\
$1$   & 1.34808 & 1.75650 & 1.3039 & 0.18913 & 1.90975 & 0.22744\\
$3/2$ & 1.20530 & 2.03672 & 1.4354 & 0.25918 & 2.21047 & 0.30262\\
$2$   & 1.11404 & 2.25316 & 1.5012 & 0.31329 & 2.36873 & 0.34218\\
$3$   & 1.03254 & 2.49075 & 1.5524 & 0.37269 & 2.52276 & 0.38069\\
$5$   & 1.00234 & 2.64373 & 1.5696 & 0.41093 & 2.64484 & 0.41121\\
$8$   & 1.00004 & 2.72004 & 1.5708 & 0.43001 & ---     & ---
\end{tabular}
\end{ruledtabular}
\end{table}

\paragraph*{Large-spin limit.} The limit in the last row of Table~1 of the main text follows directly by choosing equal intervals $\theta=\pi/2$ in the resolved protocol. The first measurement gives $q_1=+1$. A rotation by $\pi$ then takes $|j,+j\rangle$ to $|j,-j\rangle$, so $C_{13}=-1$. The spin-rotation amplitudes give $P(m_2=m)=P(m_3=-j\mid m_2=m)=p_m$, where
\begin{equation}
p_m=2^{-2j}\binom{2j}{j+m},\qquad m=-j,\ldots,j.
\end{equation}
Thus $p_{-j}=2^{-2j}$, $C_{12}=1-2p_{-j}$, and, because the resolved measurement records $m$ at $t_2$,
\begin{equation}
\begin{aligned}
C_{23}
&=1-2p_{-j}-2\sum_m p_m^2+4p_{-j}^2,\\
\sum_m p_m^2
&=2^{-4j}\binom{4j}{2j}.
\end{aligned}
\end{equation}
Consequently,
\begin{equation}
\begin{aligned}
K_1^{\mathrm R}(\pi/2)
&=3-4\,2^{-2j}-2^{1-4j}\binom{4j}{2j}+4\,2^{-4j}\\
&=3-\sqrt{\frac{2}{\pi j}}+O(j^{-1}).
\end{aligned}
\end{equation}
Here we used $\binom{4j}{2j}/2^{4j}\sim(2\pi j)^{-1/2}$. Since $K_1\le3$ for any three correlations of dichotomic outcomes, $K_{1,\max}^{\mathrm R}\to3$. Equation~\eqref{eq:Npair} then gives $\Nmin^{\mathrm R}\to(3-1)/4=1/2$.

These are maxima within one physical family, not global optima over all states, dynamics, and observables of a given dimension; global bounds for multilevel systems are discussed in Ref.~\cite{PhysRevLett.113.050401}. Our conclusions do not depend on optimality. Any observed $K$ certifies $\Nmin=[K-1]_+/4$.

\section{Multiple number of times}
\label{sec:ntimes}

For the measured cycle correlations, the exact relation is
\begin{equation}
\begin{aligned}
\Nmin^{(n)}
&=\frac{[\max_\sigma K_\sigma-(n-2)]_+}{\Delta_n},\\
\Delta_n&=
\begin{cases}
2(n-2),&n\ \text{even},\\
2(n-1),&n\ \text{odd}.
\end{cases}
\end{aligned}
\label{eq:minimum-n}
\end{equation}
Here $K_\sigma=\sum_e\sigma_eC_e$ and the cycle pairs are $(1,2),(2,3),\ldots,(n-1,n),(1,n)$. The allowed sign patterns have an odd number of minus signs. We first bound the violation using the witness spectrum, then construct weights that attain the bound, and finally prove that a positive description exists when all cycle inequalities hold.

\paragraph*{Witness spectrum.} For the chain combination $K_n=\sum_{m=1}^{n-1}C_{m,m+1}-C_{1n}$, let $s_m=q_mq_{m+1}$. Then $q_1q_n=\prod_ms_m$ and the witness value is $w=\sum_ms_m-\prod_ms_m=(n-1)-2k-(-1)^k$, where $k$ is the number of negative $s_m$. Thus $w_{\max}=n-2$ (at $k=0,1$), and $w_{\min}=-n$ for odd $n$ ($k=n-1$) and $w_{\min}=-(n-2)$ for even $n$. Every other sign pattern with an odd number of minus signs is related to the chain pattern by flipping some $q_i$, which flips the two cycle edges that meet at $i$, and so has the same spectrum. The argument of Eq.~\eqref{eq:lower} gives $K_\sigma\le w_{\max}\tr H_+-w_{\min}\tr H_-=(n-2)+\Delta_n\N(H)$ with $\Delta_n=w_{\max}-w_{\min}$.

\paragraph*{Lower and upper bounds.}
The same witness spectrum also gives a lower bound on each
LG combination. Writing $H=H_+-H_-$, with
$\tr H_+=1+\N(H)$ and $\tr H_-=\N(H)$, gives
\begin{equation}
\begin{aligned}
K_\sigma
&=\tr(H_+W_\sigma)-\tr(H_-W_\sigma)\\
&\geq w_{\min}\tr H_+-w_{\max}\tr H_-\\
&=w_{\min}-\Delta_n\N(H).
\end{aligned}
\label{eq:cycle-lower-bound}
\end{equation}
Together with the upper bound, this yields
\begin{equation}
w_{\min}-\Delta_n\N(H)
\leq K_\sigma
\leq w_{\max}+\Delta_n\N(H).
\label{eq:cycle-two-sided}
\end{equation}

For three and four measurement times, respectively, these
bounds become
\begin{align}
n=3:\quad
-3-4\N(H)&\leq K_\sigma\leq1+4\N(H),
\label{eq:three-time-two-sided}\\
n=4:\quad
-2-4\N(H)&\leq K_\sigma\leq2+4\N(H).
\label{eq:four-time-two-sided}
\end{align}
The three-time upper bound recovers Eq.~\eqref{eq:lower}. For four times, either an excess above two or a decrease below minus two therefore certifies negativity.

For even $n$, reversing every sign in an allowed pattern gives another allowed pattern, and $K_{-\sigma}=-K_\sigma$. The lower bound is therefore equivalent to the upper bound for the opposite sign pattern. Both are already included in the maximization in Eq.~\eqref{eq:minimum-n}.

For odd $n$, the lower bound does not provide an additional restriction on the measured correlations. Since each correlation lies between minus one and one, their signed sum already obeys $K_\sigma\geq-n$, which is at least as strong as the lower bound above.
\paragraph*{Minimal history weights.} Here we show that the bound $K_\sigma\le(n-2)+\Delta_n\N(H)$ is attained whenever an inequality is violated, which proves Eq.~\eqref{eq:minimum-n} in the violating region. Let $\sigma$ be the violated sign pattern, $K\equiv K_\sigma>n-2$ (at most one pattern can be violated, since two different patterns differ in at least two signs, so $K_\sigma+K_{\sigma'}\le2(n-2)$). Define the signed edge products $r_e=\sigma_e\,q_uq_v$ for each cycle pair $e=(u,v)$, and their target averages $c_e=\sigma_eC_e$, so that $\sum_ec_e=K$. Every history has $\prod_er_e=-1$, and the witness $W_\sigma=\sum_e\sigma_eQ_uQ_v$ takes the value $\sum_er_e=n-2k'$, where $k'$ is the (odd) number of $r_e=-1$. The largest value $n-2$ occurs for a single $r_e=-1$; the smallest occurs for all $r_e=-1$ (odd $n$) or for all but one $r_e=-1$ (even $n$). Set
\begin{equation}
B=\frac{K-(n-2)}{\Delta_n},\qquad s_e=\tfrac12\left(1+2B-c_e\right)\ge0 .
\end{equation}
One checks from the definition of $B$ that $\sum_es_e=1+B$ for odd $n$ and $\sum_es_e=1+2B$ for even $n$.

\emph{Odd $n$.} Give weight $s_e$ to the pattern with a single $r_e=-1$, and weight $-B$ to the pattern with all $r_e=-1$. The total weight is $\sum_es_e-B=1$, and the average of $r_f$ is $\sum_es_e-2s_f+B=1+2B-2s_f=c_f$.

\emph{Even $n$.} Give weight $a_e=s_e(1+B)/(1+2B)$ to the pattern with a single $r_e=-1$, and weight $-b_e$ with $b_e=s_eB/(1+2B)$ to the pattern in which only $r_e=+1$. Then $\sum_ea_e=1+B$ and $\sum_eb_e=B$, so the total weight is $1$, and the average of $r_f$ is $(1+B)-2a_f+B-2b_f=1+2B-2s_f=c_f$.

In both cases the weights reproduce all $n$ correlations, and the total negative weight is $B$. Each pattern of $r$'s corresponds to two histories related by the global flip; splitting its weight equally between them gives signed history weights $P(q_1,\dots,q_n)$, and the diagonal kernel $\sum_qP(q)\,|q\rangle\langle q|$ on $n$ outcome registers has $\N=B$, which meets the lower bound. If instead $\max_\sigma K_\sigma\le n-2$, a genuine probability distribution exists~\cite{Halliwell:2019uzy}; a short self-contained argument is given below. For $n=3$ the construction reproduces the zigzag weights of Eq.~\eqref{eq:zigzag} and the spectrum~\eqref{eq:pairspectrum}. We confirmed Eq.~\eqref{eq:minimum-n} for $n\le8$ by linear programming over all signed history weights.

\paragraph*{No violation implies a positive description.} Suppose $|C_e|\le1$ and $K_\sigma\le n-2$ for all $\sigma$. The set $S$ of such data is convex and compact, and $\Nmin(C)$ is a convex function (mixing optimal signed weights for two data sets gives admissible weights for the mixture). On the boundary of $S$ either (a) some $K_\sigma=n-2$, and the construction above with $B=0$ gives nonnegative weights; or (b) some $|C_e|=1$. In case (b) the two ends of $e$ are perfectly correlated or anticorrelated. After relabeling one of them if needed, merge them; the remaining data form an $(n-1)$-cycle that satisfies its own inequalities: once $C_e=+1$, any sign pattern $\sigma'$ of the $(n-1)$-cycle, extended by $\sigma_e=+1$, is a pattern $\sigma$ of the $n$-cycle, and $K'_{\sigma'}=K_\sigma-1\le(n-1)-2$. For $n=3$ the set $S$ is the tetrahedron spanned by deterministic points, so a positive distribution exists. By induction, $\Nmin=0$ on the boundary of $S$. Any interior point lies on a segment between two boundary points, so convexity and $\Nmin\ge0$ give $\Nmin=0$ inside as well. This is the correlation-level Fine theorem for cycles~\cite{BarahonaMahjoub1986,DezaLaurent1997,Halliwell:2019uzy}.

\paragraph*{Independence of the physical system.} The witness bound uses only the two outcomes of $Q$ and the commutation of observables acting on different time legs. Its eigenvalue bounds therefore do not depend on the system's dimension or dynamics.

The minimizing history weights can also be represented on physical time legs of arbitrary dimension. Let $H_{\rm out}$ be the diagonal kernel constructed above. On each physical leg, choose one normalized eigenstate for each outcome and define
\begin{equation}
V_i|q\rangle=|\phi_q^{(i)}\rangle,\qquad
Q|\phi_q^{(i)}\rangle=q|\phi_q^{(i)}\rangle,\qquad
V=\bigotimes_{i=1}^{n}V_i .
\end{equation}
The two chosen states are orthogonal, so $V$ is an isometry.
The kernel
\begin{equation}
H_{\rm phys}=V H_{\rm out}V^\dagger
\end{equation}
has the same nonzero eigenvalues as $H_{\rm out}$ and reproduces the same pair correlations. It therefore attains the same minimum negativity on the physical legs.

Neither the bound nor this construction requires the initial state, Hamiltonian or measurement update. These choices can change the measured correlations and hence the numerical minimum, but not the exact relation. Consequently, for every $n\geq3$, the least compatible negativity is nonzero if and only if at least one cycle LG inequality is violated.

\paragraph*{The pair kernel is not minimal for $n\ge4$.} On $n$ two-dimensional outcome registers, let $Q_i=\id^{\otimes(i-1)}\otimes Q\otimes\id^{\otimes(n-i)}$. The pair kernel $2^{-n}(\id^{\otimes n}+\sum_eC_eQ_{u(e)}Q_{v(e)})$ is compatible, but for $n\ge4$ its even higher moments are set to zero rather than chosen optimally. In random tests, its negativity exceeded $\Nmin^{(n)}$ in $123$ of $143$ violating samples at $n=4$ and in all $59$ violating samples at $n=5$. The four-time qubit example in the main text, $\theta=\pi/4$, is a case where it happens to be minimal.

\section{Including a second qubit observable}
\label{sec:two-observables}

The pair correlations do not determine how the system
responds to measurements of another observable.
Incompatible measurements have previously been used to
witness temporal negativity~\cite{Liu:2025dzl}.
Here we determine the exact minimum negativity required
by five specified averages and construct a kernel
attaining it.

\paragraph*{Measured averages.}
In this section the time legs are copies of the physical
qubit space. They are not registers recording two
independent classical outcomes. Choose
\begin{equation}
Q=\sigma_z,\qquad
S=\sigma_x,\qquad
Q^2=S^2=\id,\qquad
QS+SQ=0.
\label{eq:QS-algebra}
\end{equation}
In addition to the three usual pair correlations,
measure $S$ only at the middle time in one set of runs.
In another set, measure $Q,S,Q$ sequentially using the
L\"uders rule. All runs use the same initial preparation.
The two additional averages are
\begin{equation}
\begin{aligned}
m_S
&=\langle s_2\rangle
=\tr[\rho_0S(t_2)],\\
T_S
&=\langle q_1s_2q_3\rangle\\
&=\frac14\tr\!\left[
\rho_0
\{\{Q(t_3),S(t_2)\},Q(t_1)\}
\right].
\end{aligned}
\label{eq:additional-averages}
\end{equation}
The second identity follows from the weighted L\"uders
update in Eq.~\eqref{eq:jordan}, or equivalently from
Eq.~\eqref{eq:F} with the middle observable replaced by $S$.

Write $S_2=\id\otimes S\otimes\id$, with the leg
observables $Q_i$ defined as before. We minimize over
operators satisfying
\begin{equation}
\begin{aligned}
H&=H^\dagger,\qquad \tr H=1,\\
\tr(HQ_iQ_j)&=C_{ij},\qquad i<j,\\
\tr(HS_2)&=m_S,\qquad
\tr(HQ_1S_2Q_3)=T_S.
\end{aligned}
\label{eq:five-constraints}
\end{equation}
As in the pair-only minimum, no additional requirement
is imposed that each compatible operator arise from a
particular Hamiltonian or measurement process.

\paragraph*{An optimal kernel.}
Consider
\begin{equation}
\begin{aligned}
H_\star=\frac18\Big[
&\id^{\otimes3}
+C_{12}(Q\otimes Q\otimes\id)
+C_{23}(\id\otimes Q\otimes Q)\\
&+C_{13}(Q\otimes\id\otimes Q)
+m_S(\id\otimes S\otimes\id)
+T_S(Q\otimes S\otimes Q)
\Big].
\end{aligned}
\label{eq:five-kernel}
\end{equation}
Orthogonality of the Pauli products shows that this
operator satisfies Eq.~\eqref{eq:five-constraints}.
We next prove that no compatible operator has less
negativity.

Start with an arbitrary compatible $H$ and average over
conjugation by $Q$ on the two outer legs,
\begin{equation}
H^{(1)}
=\frac14\sum_{a,b=0}^{1}
Q_1^aQ_3^bH Q_3^bQ_1^a.
\label{eq:outer-average}
\end{equation}
Since conjugation by $Q$ reverses the signs of
$\sigma_x$ and $\sigma_y$, all terms containing either
of these matrices on an outer leg cancel. Only $\id$
and $Q$ remain on those legs.

Now define $\mathcal F=S\otimes S\otimes S$ and set
\begin{equation}
H^{(2)}
=\frac12\left[
H^{(1)}+\mathcal F H^{(1)}\mathcal F^\dagger
\right].
\label{eq:second-average}
\end{equation}
Conjugation by $S$ reverses the signs of $Q$ and
$Y=\sigma_y$ while preserving $\id$ and $S$.
Consequently, only terms with an even total number of
$Q$ and $Y$ factors survive. The remaining operator has
the form
\begin{equation}
H^{(2)}
=H_\star+\frac18\left[
u(Q\otimes Y\otimes\id)
+v(\id\otimes Y\otimes Q)
\right],
\end{equation}
where $u$ and $v$ are real coefficients not fixed by
the five measurements.

In the basis where $Q=\sigma_z$, transposition preserves
$\id,Q,S$ and reverses the sign of $Y$. Thus
\begin{equation}
\frac12\left[
H^{(2)}+(H^{(2)})^{\mathsf T}
\right]=H_\star.
\label{eq:transpose-average}
\end{equation}
All three averages preserve the five constraints.
Unitary conjugation and transposition preserve the trace
norm, whereas averaging cannot increase it. For example,
\begin{equation}
\left\|
\frac{H^{(2)}+(H^{(2)})^{\mathsf T}}2
\right\|_1
\leq
\frac{
\|H^{(2)}\|_1+
\|(H^{(2)})^{\mathsf T}\|_1
}{2}
=\|H^{(2)}\|_1.
\end{equation}
Applying the same argument to the other averages gives
\begin{equation}
\|H_\star\|_1
\leq\|H^{(2)}\|_1
\leq\|H^{(1)}\|_1
\leq\|H\|_1.
\end{equation}
Every operator retains unit trace, so
\begin{equation}
\N(H_\star)\leq\N(H).
\end{equation}
Since $H_\star$ is itself compatible, it attains the
minimum. Transposition is used here only as a step in
the proof, not as an operation required in the experiment.

\paragraph*{Spectrum and minimum negativity.}
Fix the outer outcomes $q_1,q_3=\pm1$. On the middle
qubit, Eq.~\eqref{eq:five-kernel} becomes
\begin{equation}
B(q_1,q_3)=\frac18\Big[
(1+q_1q_3C_{13})\id
+(q_1C_{12}+q_3C_{23})Q
+(m_S+q_1q_3T_S)S
\Big].
\end{equation}
Set $p=q_1q_3$ and define
\begin{equation}
a_p=C_{12}+pC_{23},\qquad
b_p=m_S+pT_S.
\end{equation}
Since $q_3=pq_1$, the matrix is
\begin{equation}
B(q_1,pq_1)
=\frac18
\begin{pmatrix}
1+pC_{13}+q_1a_p & b_p\\
b_p & 1+pC_{13}-q_1a_p
\end{pmatrix}.
\end{equation}
Its characteristic equation is
\begin{equation}
(1+pC_{13}-8\lambda)^2-a_p^2-b_p^2=0.
\end{equation}
Therefore
\begin{equation}
\lambda_{p,\pm}
=\frac{1+pC_{13}\pm R_p}{8},
\qquad
R_p=\sqrt{(C_{12}+pC_{23})^2+(m_S+pT_S)^2}.
\label{eq:five-spectrum}
\end{equation}
For each $p$, the two choices of $q_1$ give the same
eigenvalues. Each eigenvalue in Eq.~\eqref{eq:five-spectrum}
thus occurs twice.

Because $|C_{13}|\leq1$, the plus branch is nonnegative. The minus branch is negative precisely when $R_p>1+pC_{13}$. Its two copies then contribute $(R_p-1-pC_{13})/4$. Hence
\begin{equation}
\Nmin^{Q,S}
=\N(H_\star)
=\frac14\sum_{p=\pm1}
\left[
\sqrt{(C_{12}+pC_{23})^2+(m_S+pT_S)^2}
-1-pC_{13}
\right]_+ .
\label{eq:five-minimum}
\end{equation}
In particular, the five averages admit a positive kernel on the qubit time legs if and only if
\begin{equation}
(C_{12}+pC_{23})^2+(m_S+pT_S)^2
\leq(1+pC_{13})^2
\qquad\text{for both }p=\pm1.
\label{eq:five-positivity}
\end{equation}

When the additional averages vanish,
\begin{align}
\left.\Nmin^{Q,S}\right|_{m_S=T_S=0}
&=\frac14\Big(
[|C_{12}+C_{23}|-1-C_{13}]_+
+[|C_{12}-C_{23}|-1+C_{13}]_+
\Big)
=\Nmin.
\end{align}
where the last equality follows from Eq.~\eqref{eq:Npair}. The result therefore extends the pair-only minimum rather than replacing it.

\paragraph*{Linear witness inequalities.}
The same five averages define a family of measurable combinations,
\begin{equation}
\mathcal L_{p,\phi}
=\cos\phi(C_{12}+pC_{23})
+\sin\phi(m_S+pT_S)
-pC_{13},
\qquad p=\pm1.
\label{eq:two-observable-combination}
\end{equation}
Their witness operators are
\begin{equation}
\begin{aligned}
\mathcal W_{p,\phi}
={}&\cos\phi(Q_1Q_2+pQ_2Q_3)\\
&+\sin\phi(S_2+pQ_1S_2Q_3)-pQ_1Q_3.
\end{aligned}
\end{equation}
To find the spectrum, fix the outer product $r=q_1q_3$. If $r=-p$, the witness equals the identity on the middle qubit. If $r=p$, it becomes
\begin{equation}
-\id+2\left(q_1\cos\phi\,Q+\sin\phi\,S\right).
\end{equation}
The anticommutation relation gives
\begin{equation}
\left(q_1\cos\phi\,Q+\sin\phi\,S\right)^2=\id.
\end{equation}
Its eigenvalues are therefore $\pm1$, so every $\mathcal W_{p,\phi}$ has only the eigenvalues
$1$ and $-3$. The argument used in Eq.~\eqref{eq:lower} consequently gives
\begin{equation}
\mathcal L_{p,\phi}\leq1+4\N(H).
\label{eq:two-observable-bound}
\end{equation}
Every positive kernel obeys the bound with right-hand side one.

Optimizing over the angle gives
\begin{equation}
\max_\phi\mathcal L_{p,\phi}=R_p-pC_{13}.
\end{equation}
Together with Eq.~\eqref{eq:five-minimum}, this proves
\begin{equation}
\Nmin^{Q,S}>0
\quad\Longleftrightarrow\quad
\mathcal L_{p,\phi}>1
\quad\text{for some }p,\phi.
\label{eq:two-observable-iff}
\end{equation}
Both values of $p$ can contribute to the minimum.
Unlike the four ordinary LG combinations, their
optimized violations need not be mutually exclusive.

At angles zero and $\pi$, the family reduces to the
four ordinary LG combinations,
\begin{equation}
\mathcal L_{+,0}=K_1,\qquad
\mathcal L_{-,0}=K_2,\qquad
\mathcal L_{-,\pi}=K_3,\qquad
\mathcal L_{+,\pi}=K_4.
\end{equation}
For general angles, however, the bound uses the known
qubit relation between $Q$ and $S$. It is a condition
on positive operator descriptions, not a classical
LG inequality derived solely from assigning independent
signs to the two observables.

\paragraph*{Precessing-qubit example.}
Use the Hamiltonian and equal intervals of Sec.~\ref{sec:qubit}, with the first time set to zero. Since the Hamiltonian is proportional to $\sigma_x$,
\begin{equation}
S(t)=\sigma_x,\qquad
Q(t)=\sigma_z\cos\Omega t+\sigma_y\sin\Omega t.
\end{equation}
Both terms in $Q(t)$ anticommute with $\sigma_x$. Consequently,
\begin{equation}
\{Q(t_3),S(t_2)\}=0,\qquad
T_S=0,\qquad
m_S=\tr(\rho_0\sigma_x).
\end{equation}

Prepare the state
\begin{equation}
\rho_0=\frac{\id+\sigma_x}{2}
=|\psi_x\rangle\langle\psi_x|,
\qquad
|\psi_x\rangle
=\frac{|0\rangle+|1\rangle}{\sqrt2}.
\end{equation}
Here $|0\rangle$ and $|1\rangle$ are the positive and negative eigenstates of $Q=\sigma_z$. Measuring $S$ immediately gives plus one with certainty. Thus $m_S=1$.

At $\theta=\pi/3$, the five averages are
\begin{equation}
(C_{12},C_{23},C_{13})
=\left(\frac12,\frac12,-\frac12\right),
\qquad m_S=1,\qquad T_S=0.
\end{equation}
The radii entering Eq.~\eqref{eq:five-minimum} are
\begin{equation}
R_+=\sqrt2,\qquad R_-=1,
\end{equation}
whereas the corresponding positive-kernel bounds are
\begin{equation}
1+C_{13}=\frac12,\qquad
1-C_{13}=\frac32.
\end{equation}
Therefore
\begin{equation}
\Nmin^{Q,S}
=\frac14\left(
[\sqrt2-\tfrac12]_+
+[1-\tfrac32]_+
\right)
=\frac{2\sqrt2-1}{8}
\simeq0.2286.
\end{equation}
The witness with $p=+1$ and $\phi=\pi/4$ gives
\begin{equation}
\mathcal L_{+,\pi/4}
=\sqrt2+\frac12.
\end{equation}
Its violation in Eq.~\eqref{eq:two-observable-bound}
certifies exactly the same minimum.

The ordinary LG result is unchanged,
\begin{equation}
K_1=\frac32,\qquad \Nmin=\frac18.
\end{equation}
For this initial state all single-time averages of $Q$ vanish, because $Q(t)$ lies in the $y$--$z$ plane. Equation~\eqref{eq:qubitodd} then also gives a vanishing three-time correlation of $Q$. Hence
\begin{equation}
H_Q=\Hpair,\qquad
\N(H_Q)=\frac18<\Nmin^{Q,S}.
\end{equation}
The second observable therefore certifies negativity that even a complete description of the histories of $Q$ alone does not require.

For comparison, the initial state $|0\rangle$ gives the same three LG pair correlations, but both added averages vanish,
\begin{equation}
m_S=0,\qquad T_S=0,\qquad
\Nmin^{Q,S}=\frac18.
\end{equation}
Thus identical LG violations need not give identical certificates once another observable is included.

The additional inequalities can also be violated when all four ordinary LG inequalities hold. With the $\rho_0=(\id+\sigma_x)/2$ preparation and $\theta=\pi/2$,
\begin{equation}
\begin{aligned}
(C_{12},C_{23},C_{13})&=(0,0,-1),\\
(K_1,K_2,K_3,K_4)&=(1,-1,-1,1),\\
m_S&=1,\qquad T_S=0.
\end{aligned}
\end{equation}
Nevertheless,
\begin{equation}
\mathcal L_{+,\pi/2}=m_S+T_S-C_{13}=2,
\qquad
\Nmin^{Q,S}=\frac14.
\end{equation}

\paragraph*{The bound from the full kernel.}
Let $\mathcal A$ denote a set of measured operators, all acting on the same time-leg space as $H^{\mathrm L}$. Their averages are
\begin{equation}
c_A=\tr(H^{\mathrm L}A),
\qquad A\in\mathcal A.
\end{equation}
Define $\Nmin(\mathcal A)$ by minimizing over Hermitian, unit-trace kernels reproducing all these averages. The full kernel is one allowed choice, so
\begin{equation}
\Nmin(\mathcal A)\leq\N(H^{\mathrm L}).
\label{eq:full-kernel-ceiling}
\end{equation}
If $\mathcal A\subseteq\mathcal B$, reproducing the averages in $\mathcal B$ imposes every constraint in $\mathcal A$ and possibly more. The set over which we minimize can only become smaller. Therefore
\begin{equation}
\Nmin(\mathcal A)
\leq\Nmin(\mathcal B)
\leq\N(H^{\mathrm L}).
\end{equation}
If the measured operators span the full operator space, their averages determine $H^{\mathrm L}$ uniquely and the last inequality becomes equality.

Accordingly, $\N(H^{\mathrm L})$ is the largest value that complete measurements can certify for this fixed process. It is not the largest negativity among arbitrary kernels reproducing an incomplete set of averages. Nor does the argument imply that deleting arbitrary terms from $H^{\mathrm L}$ decreases its negativity.

For the closed-qubit, three-time L\"uders construction, Eq.~\eqref{eq:HLspec} also gives a maximum over physical initial states. The construction is linear in the state, and negativity is convex. For any decomposition
\begin{equation}
\rho_0=\sum_a w_a|\psi_a\rangle\langle\psi_a|,
\qquad w_a\geq0,\qquad \sum_a w_a=1,
\end{equation}
we have
\begin{equation}
\begin{aligned}
\N(H^{\mathrm L}[\rho_0])
&\leq
\sum_a w_a
\N(H^{\mathrm L}[|\psi_a\rangle\langle\psi_a|])\\
&=\frac{1+\sqrt2+\sqrt3}{4}.
\end{aligned}
\end{equation}
Every pure initial state attains this value, and unitary evolution preserves it by Eq.~\eqref{eq:covariance}. Thus, within this stated family,
\begin{equation}
\Nmin^{Q,S}
\leq\N(H^{\mathrm L})
\leq\frac{1+\sqrt2+\sqrt3}{4}.
\end{equation}
This last numerical ceiling is specific to three complete qubit time legs with unitary evolution. It is not a dimension-independent bound.
\bibliography{ref}

\end{document}